\def\etal{{\it et al.\/}}

\def\ie{{\it i.e.\/}}

\documentstyle[aaspp]{article}
\tighten
\singlespace

\begin{document}

\title{The afterglow of gamma ray bursts II: \\
the cases of GRB 970228 and GRB 970508}
\author{Mario Vietri}
\affil{
Dipartimento di Fisica E. Amaldi, Universit\`a di Roma 3, \\
via della Vasca Navale 84, 00147 Roma, Italy \\
E--mail: vietri@corelli.fis.uniroma3.it \\
}

\begin{abstract}
Highly radiative expansion of a relativistic shell is shown to 
explain all observed features of the afterglows of the two bursts
{\it GRB 970228} and {\it GRB 970508}. In particular, in the
first case the observed time--dependence $\propto t^{-1.32}$ of the soft 
X--ray flux is easily reproduced. The same model, when the surrounding matter
density scales as a $r^{-2}$, explains the afterglow of {\it GRB 970508}, 
which may at first sight appear at odds with that of {\it GRB 970228}. 
In particular, it is shown that both the late peak in the optical luminosity 
and the flat time--dependence of the X--ray luminosity are simultaneously
explained by nonuniformity of the surrounding matter, that the observed 
optical time--delay is correctly reproduced for standard parameter values, and 
that the time--delay and flux levels of the radio emission are also explained.

\end{abstract}

\section{Introduction}

It is an easy prediction of the fireball model that gamma ray bursts (GRBs)
should show an afterglow, in the X--ray (Vietri 1997, Paper I)  and
at optical/radio wavelengths (M\`esz\`aros and Rees 1997). The energy source
powering this luminosity is the kinetic energy of the (burn out) ejecta
shell after the burst. This after--glow has become detectable thanks to 
an exceptional effort by the BeppoSAX science team which has managed
to carry out TOOs just $8$ hours after the burst {\it GRB 970228}, and then
again about $48$ hours later, thus providing the most complete coverage
of the post-burst evolution of a GRB ever obtained in the X-ray. 

By combining
observations of the Narrow Field Instruments (those of the TOOs) with both
ASCA and Wide Field Cameras observations, Costa \etal\/ (1997a) were able 
to identify the main features of {\it GRB 970228} as follows. First,
the major burst lasts only $\approx 5 s$; the afterglow begins just $30 s$ 
after the major burst, has a time--dependence $\propto 
t^{-\delta}$, with $\delta = 1.32 \pm 0.19$, and the total fluence in the 
afterglow, as measured in just the $2-10\; keV$ band, equals $40\%$ of the 
whole burst (including the $\gamma$--ray fluence!).

I will show in the following that this behaviour can be 
easily understood within the model of Paper I. In particular, I shall
show that the afterglow is due to the expansion of a
highly radiative relativistic shell of matter, expanding in a constant
density environment in the case of {\it GRB 970228}.

A considerably different time behaviour is displayed by {\it GRB 970508}
(Costa \etal, 1997b)
which shows an X--ray luminosity that decays very slowly (Costa,
talk delivered at the Elba Workshop on GRBs; Piro \etal, 1997), 
and simultaneously an optical transient peaking well after the burst 
(Djorgovski \etal, 1997b). 
Qualitatively this behaviour is easy to understand. Since
in the afterglow the dominant emission mechanism, synchrotron, is 
negligible for frequencies smaller than $\nu_m$, the synchrotron
turnover frequency, for the optical
emission to set in one has to wait for relativistic effects to allow
$\nu_m$ to move into the optical region. At the same time, the X--ray
luminosity, which is well beyond $\nu_m$, is slow in decreasing. It thus
appears as if the ejecta shell manages to remain relativistic longer
than for expansion in a uniform medium ({\it GRB 970228}); expansion in
a surrounding medium with decreasing density will be shown to provide a
quantitative, as well as qualitative, explanation for the afterglow of
{\it GRB 970508}. 

Understanding the behaviour of {\it GRB 970508}
seems especially important, in view of the minimum redshift determination
of the optical transient (Metzger \etal\/ 1997), which establishes
the cosmological nature of GRBs: so long as ROSAT HRI observations
linking precisely the site of the optical and X--ray transients are not
forthcoming, even theoretical arguments linking the two may be valuable.

In the next Section, I will discuss qualitative features in {\it GRB 970228}
which support the fireball model, while the hydrodynamics and X--ray emission 
will be discussed in Section 3. The optical and radio emissions are then 
discussed in Section 4.

\section{Qualitative deductions}

The light curve of {\it GRB 970228} in both the $\gamma$ and $X$--ray
bands is peculiar, in that it clearly shows a first peak lasting about 
$\approx 5\; s$, a silence of $\approx 30\; s$, and a new, longer peak lasting
$\approx 40 \; s$. Costa \etal\/ (1997a) attributed the second peak
to the afterglow. There are three reasons for setting the second peak
apart from the first one: first, 
the $\gamma/X$ ratio is radically different in the two events. Second, 
the luminosity of the second peak falls squarely on the extrapolation to 
early times of the time-dependence law that links the first and second TOOs 
to ASCA data. Third, the X and $\gamma$ light curves in the second peak
start out nearly simultaneously, suggesting a common origin in a new shock, 
rather than the beginning of cooling in hot matter.
Recognizing the second peak as part of the afterglow implies that
the observed afterglow follows accurately a power--law time--dependence
over nearly four orders of magnitude in time and five in flux, without
apparently any band effect. 

The total soft X--ray ($2-10\; keV$) luminosity released in the afterglow
can then be integrated to show that it alone equals $40\%$ of the whole
burst luminosity (Costa \etal\/ 1997a). Since the photon number counts go as 
$\propto \nu^{-1.9}$, the total correction to obtain the bolometric 
luminosity must amount to a factor of a few, without changing the order of
magnitude of the afterglow fluence. In short, the afterglow radiates 
about as much as the burst. 

This fact has two important consequences: first, the total energy radiated away
by the expanding fireball is large. Thus, the expansion following the 
burst cannot be described by an adiabatic expansion, as was implicitly
or explicitly assumed by other authors (Tavani 1997, Waxman 1997a, 
Wijers, Rees and M\`esz\`aros 1997), who treated radiative losses as a
tiny perturbation to an otherwise adiabatic flow. Quantitative details 
pertaining to the radiative impulsive relativistic expansion will be derived
in the next Section.

Second, the approximate equality of fluences in the burst and in the afterglow
makes it likely that the
energy source powering the burst (the kinetic energy of the ejecta) is
the same as that of the afterglow: if it were otherwise, \ie\/ if the 
the burst and afterglow were powered by different physical phenomena,
an unlikely coincidence would result. 

This equality is naturally accounted for in fireball models: when the 
Lorenz factor of the bulk expansion is low ($\eta \approx 100$), the
reverse shock is only marginally relativistic, and the total directed
kinetic energy converted into internal energy, to be radiated as the
burst, is $1/2$ of the total energy budget (Sari and Piran 1995). When,
viceversa, $\eta$ is large, $\approx 1000$, the reverse shock is highly
relativistic, and the ejcta are stopped dead in their tracks. In this case,
the energy fraction converted into internal energy $\approx 1$, and the
balance remaining as directed kinetic energy $\ll 1$, leading to a
much weaker afterglow. This difference may account for the lack of afterglow
from {\it GRB 970111}, or its relative weakness in {\it GRB 970402}.

The following quantitative argument, supporting $\eta \approx 100$, 
can be made. Of the two distinct mechanisms proposed to explain GRBs,
for this low value of $\eta$ only internal shocks
(M\`esz\`aros and Rees 1994) can work,
the other one, external shocks (Rees and M\`esz\`aros 1994) requiring
much higher Lorenz factors $\eta \approx 1000$. In the interior case, two
shells collide and stick at $r\approx 10^{12}\; cm$, and from then on 
they continue a free expansion until they impinge upon the ISM, producing
a second shock. This occurs at a radius 
\begin{equation}
\label{shock}
r_{sh} = \left(\frac{3 E}{4\pi n_1 m_p c^2 \eta^2}\right)^{1/3} =
2.5\times 10^{16}\; cm\; E_{51}^{1/3} n_1^{-1/3}
\left(\frac{100}{\eta}\right)^{2/3}\;,
\end{equation}
where the total energy release is $E= E_{51} 10^{51}\; erg$, and the
ISM mass density is $n_1 m_p \;cm^{-3}$. 
Thus, as seen from Earth the time--delay between the first and second shock
and ensuing emissions is $d\!t \approx r_{sh}/2\eta^2 c = 40\;s (\eta/100)^
{-8/3}$. This value is essentially the time--delay between the first
peak of the {\it GRB 970228}, and the onset of the afterglow. Also, 
the simultaneity of the onset of X--ray and $\gamma$--ray luminositites
argues for a common origin in a new shock. I thus find the internal/external
shock model for GRBs validated by the observed time--delay in {\it GRB 970228},
and the value of $\eta \approx 100$ supported by observations.

\section{The X--ray luminosity}

The time--dependence
of the soft X--ray flux in a highly relativistic, radiative snowplow
model was derived in paper I for late times ($\approx 1$ month after the
burst). I now derive the prediction for early post--burst times. 

The initial burst is contaminated by a baryon mass $M_b = E/\eta c^2$ with
$\eta \approx 100$; the shock with the ISM occurs at a radius given by Eq. 
\ref{shock}. After this shock, the ejecta keep plowing through the ISM, 
shocking ISM matter and transforming its directed kinetic energy into internal 
energy. Assuming the post--shock cooling time to be short (to be checked later),
the shocked matter piles up behind the shock in a cold shell, whose Lorenz 
factor $\gamma$, for $\gamma \gg 1$,
evolves according to (Blandford and McKee 1976) $d\!\gamma/\gamma^2 
= -d\!M /M$, where $M$ is the total mass entrained by the shock, including
the initial contamination. Assuming the surrounding matter density to vary
with distance from the site of energy release $r$ as $\rho \propto
r^{-\alpha}$, it can easily be shown that 
\begin{equation}
\label{gamma}
\gamma = \eta \left(\frac{r_{sh}}{r}\right)^{3-\alpha}
\end{equation}
and the time, as seen from Earth $t_E$, which scales as $d\!t_E = dt/2\gamma^2
= d\!r/2\gamma^2 c$, varies as 
\begin{equation}
\label{tr}
t_E = \frac{r_{sh}}{(7-2\alpha) 2 \eta^2 c} \left( 
\left(\frac{r}{r_{sh}}\right)^ {7-2\alpha} - 1 \right)
\end{equation}
which, together, with Eq. \ref{gamma} yields
\begin{equation}
\label{tempo}
\gamma = \eta \left(\frac{t_E}{t_\circ} + 1\right)
^{-\frac{3-\alpha}{7-2\alpha}}\;,
\end{equation}
where of course
\begin{equation}
\label{t0}
t_\circ \equiv \frac{r_{sh}}{(7-2\alpha) 2 \eta^2 c} =
\frac{42\; s}{7-2\alpha} \left(\frac{100}{\eta}\right)^{8/3}\;.
\end{equation}
The time $t_\circ$ is the time--scale on which the scale--free solution 
sets in. Since we are mainly interested in times $t \gg t_\circ$, I shall from
now on neglect the $\pm 1$ arguments inside the parentheses in Eqs. \ref{tr} 
and \ref{tempo}. 

The parameter $\alpha$ can also be thought of as a ruse: for $\alpha=3/2$,
we recover the results for the adiabatic shell. It should also be noticed
that $\alpha$ is directly observable, as will be shown in the discussion 
following Eq. \ref{turnover}.  

The above solution is correct only for relativistic expansion; it breaks
down for $\gamma \approx 1$, \ie, at a time $t_r$ 
\begin{equation}
t_r = \eta^{\frac{7-2\alpha}{3-\alpha}} t_\circ = 
\frac{9\times 10^6\;s}{7-2\alpha} \eta^{\frac{7-2\alpha}{3-\alpha}-\frac{8}{3}} 
\end{equation}
which is very weakly dependent on $\alpha$. The total time through which 
the shell remains relativistic is thus about a month, as shown in Paper I.

The copious fluence in the afterglow phase (Costa \etal, 1997a) is evidence
{\it per se} that the expansion must be highly radiative; however, this
can also be checked through a simple, microphysical argument. 
For synchrotron emission, the synchrotron cooling time
in the shell frame is $t_s = 6\pi m_e c/\sigma_T \gamma_e B^2$, to be
compared with the local expansion time scale $t_{exp}= r/\gamma c$. I find
\begin{equation}
\label{ratio}
{\cal R} \equiv \frac{t_s}{t_{exp}} = \frac{3}{4} \left(\frac{m_e}{m_p}\right)^2
\frac{1}{\sigma_T n_1 r_{sh} \gamma} \frac{r_{sh}}{r}
\end{equation}
where I used an equipartition magnetic field $B =(8\pi n_1 m_p c^2)^{1/2} 
\gamma$, and energy equipartition between electrons and protons, $\gamma_e
\approx m_p \gamma/m_e$. Using Eqs. \ref{shock} and \ref{gamma}, I find
\begin{equation}
{\cal R} = 0.4 \left(\frac{\gamma}{\eta}\right)
^{1/(3-\alpha) -1} \frac{1}{\eta^{1/3}}
\end{equation}
for a typical explosion, $E_{51} = 4$ (Piran 1992). It can thus be seen
that, for $\alpha=2$, ${\cal R} \ll 1$, while, for $\alpha = 0$,
${\cal R} = 1$ for $\gamma = 2.8$, close enough to the limit of validity
($\gamma \gg 1$) of Eq. \ref{tempo} to allow us to say that ${\cal R} < 1$
is verified through the whole period in which the shell is relativistic.

In the highly radiative limit, the total power radiated per unit time
is given by a purely hydrodynamical argument (Blandford and McKee 1976).
We have $dE/dt= 4\pi r^2c \gamma^2 n_1 m_p c^2$, in the relativistic 
limit $\gamma \gg 1$; in Earth time $dE/dt_E= 2\gamma^2 dE/dt$. I find 
$\frac{dE}{dt_E} \propto t_E^{\frac{4\alpha-10}{7-2\alpha}}$.

The fraction of all power radiated in the X--ray region, $f_X$, is
computed in Paper I as $f_X = (\epsilon_u^{(3-p)/2}-\epsilon_l^{(3-p)/2})
/\epsilon_m^{(3-p)/2}$; here the index $p=2.8$ is the spectral index
of the electron power--law energy distribution, $d\!n \propto \gamma_e^{-p}$,
with $p$ fixed by the requirement that the observed spectrum, $\propto
\nu^{-1.9}$ (Costa \etal, 1997a) be reproduced. Also, $\epsilon_u$ and
$\epsilon_l$ are the upper and lower limits of the BeppoSAX instruments,
$2$ and $10$~keV respectively; $\epsilon_m$ is the maximum electron energy.
Proceeding as in Paper I, I find $f_X = (6\times 10^{-3}/\gamma)^{(3-p)/2}$.
In the end, I find
\begin{equation}
\label{xflux}
F_X = 4.8\times 10^{-6} \; erg\; s^{-1}\; cm^{-2} \left(\frac{t_E}{1\; s}
\right)^{-\delta}
\end{equation}
where 
\begin{equation}
\delta = \frac{10-4\alpha}{7-2\alpha} + \frac{3-p}{2}\frac{3-\alpha}
{7-2\alpha}\;.
\end{equation}
The constant above is computed for $\alpha=0$, and a source distance
$D=2 \; Gpc$. It varies little with $\alpha$, and given our ignorance
of the source distance, it is not worth it to give its (complex)
dependence on other parameters. Since $3-p \ll 1$, the dependence
upon waveband is very weak, explaining why SAX can observe such a
strikingly extended power--law. 

The above discussion allows an immediate fit to the properties of 
{\it GRB 970228}. In this case, $p=2.8$, and $\alpha=0$ (constant
density) imply $\delta = 1.38$, in excellent agreement with the
observed $\delta = 1.32 \pm 0.19$. This is clearly displayed in
Fig. 1. {\it GRB 970508} has a different time--behaviour, showing only
a modest decline in flux by a factor $4.5$ from $0.6d$ after the burst
to $6.1 d$ after the burst (Piro \etal, 1997); this corresponds to an average 
decrease going like $F_X \propto t^{-0.65}$. Since the spectral properties 
of {\it GRB 970228} and {\it GRB 970508} seem similar, we can
use the equation above to see that, for $\alpha=2$, we have 
$F_X \propto t^{-0.63}$, in excellent agreement with observational 
data. It should be noticed that these data require a density gradient:
for the highly radiative evolution, it was shown above that expansion
in a constant density environment leads to $F_X \propto t^{-1.38}$, 
while adiabatic evolution leads to $F_X \propto t^{-3z/2}$, where
$z \approx 0.9$, the spectral slope of the burst, is most likely
similar for {\it GRB 970228} and {\it GRB 970508}.

\section{The optical and radio emission}

One may think at this point that the introduction of a declining density
into which the fireball plows is but a simple minded way to account
for the slowly declining X--ray flux, but I show here that this stratagem 
simultaneously explains why the optical luminosity is seen increasing,
yielding quantitative predictions in excellent agreement with observations. 

Typical spectra of GRBs are of the form $I_\nu \propto \nu^a$
for $\nu < \nu_m$, and $I_\nu \propto \nu^b$ for $\nu > \nu_m$,
with $a \approx 0$ but very ill--determined, and $b\approx -1$. 
Since the equipartition
magnetic field $B \propto \gamma$, the turnover frequency as seen
from the Earth $\nu_m \propto \gamma B \gamma_m^2$, where $\gamma_m 
\propto \gamma$, so that $\nu_m \propto \gamma^4$. The electron
density in the shell frame $n_e \propto \gamma$, and the comoving
shell thickness $d\!r = r/\gamma$, so that the comoving intensity
$I_{\nu_m} \propto n_e B^2 \gamma_m^2 d\!r/ B \gamma_m^2 \propto 
n_e B d\!r$, and the observed flux as a function of observer time
is $F_{\nu_m} \propto t^2 \gamma^5 I_{\nu_m} \propto t^2 \gamma^6
r$ (Wijers, Rees and M\`esz\`aros 1997). From the above Eqs. \ref{tr} and 
\ref{tempo} I find $F_{\nu_m} \propto t^q$, where 
$q=(2\alpha-3)/(7-2\alpha)$.
So long as $\nu_m$ is shortward of the optical region, the
optical luminosity is $\propto F_{\nu_m} (\nu_{opt}/\nu_m)^a 
\propto t^p$, where $p = q+4a(3-\alpha)/(3(7-2\alpha))$.
For $a\approx 0$ and $\alpha=2$, I find $p\approx q \approx 1/3$, 
but ill--determined because of the uncertainty on $a$. 

Once $\nu_m$ has entered the optical region, the luminosity must then 
decrease according to 
\begin{equation}
\label{optflux}
F_{opt} = F_{\nu_m} (\nu_{opt}/\nu_m)^b \propto t^p \;\;;\;\;
p=(2\alpha-3)/(7-2\alpha) +4 b \frac{3-\alpha}{7-2\alpha}\;.
\end{equation}
For $\alpha=2$, $p= -1$, for $b=-1$. The optical data from Table I
are shown in Fig. 2, corrected for standard Galactic absorption and
for a spectral shape $\propto \nu^{-1}$. They show clearly that both 
the rise (less significantly) and the decline (more significantly)
agree with the model for $\alpha =2$. 

\begin{table}
\begin{center}
   \caption[]{Optical fluxes of {\it GRB 970508}. Time is
              measured from burst trigger, May 8.904 UT (Costa \etal, 1997b). 
              }
   \label{ta:flux}
\begin{tabular}{llllll} \hline
$\log t$ & app. magn.  & $\log\nu$ & band & instrument & ref. \\
 (s)     & $m$         & (Hz)      &      &            &       \\ \hline
 4.39 &$\approx$21.5 &  14.74 & V & Kitt Peak & 1 \\
 5.04 & 20.5 & 14.74 & V & Kitt Peak & 1 \\
 5.04 & 19.6 & 14.56 & I & Kitt Peak & 1 \\
 4.41 & 21.2 & 14.64 & R & Kitt Peak & 2 \\
 5.00 & 21.5 & 14.92 & U & La Palma & 2 \\
 5.00 & 21.0 & 14.74 & V & La Palma & 2 \\
 5.00 & 20.35 & 14.64 & R & La Palma & 2 \\
 5.00 & 20.2 & 14.56 & I & La Palma & 2,7 \\
 5.26 & 20.5 & 14.92 & U & La Palma & 2 \\
 5.26 & 20.3 & 14.84 & B & La Palma & 2,7 \\
 5.26 & 20.2 & 14.74 & V & La Palma & 2 \\
 5.26 & 20.1 & 14.64 & R & La Palma & 2 \\
 5.26 & 19.1 & 14.56 & I & La Palma & 2,7 \\
 4.39 & 21.33 & 14.65 & Gunn-r & Palomar & 3 \\
 5.06 & 20.17 & 14.65 & Gunn-r & Palomar & 3 \\
 5.30 & 20.15 & 14.65 & Gunn-r & Palomar & 3 \\
 5.24 & 19.65 & 14.74 & V & NOT & 4 \\
 5.45 & 20.53 & 14.65 & Gunn-r & Palomar & 5 \\
 5.57 & 20.76 & 14.65 & Gunn-r & Palomar & 6 \\
 5.43 & 20.9 & 14.92 & U & La Palma & 7 \\
 5.43 & 20.9 & 14.84 & B & La Palma & 7 \\
 5.43 & 20.6 & 14.74 & V & La Palma & 7 \\
 5.43 & 20.2 & 14.64 & R & La Palma & 7 \\
 5.22 & 19.8 & 14.64 & R & Loiano & 8 \\
 5.46 & 20.47 & 14.64 & R & M. Hopkins & 9 \\ 
 6.33 & 23.10 & 14.64 & R & HST & 10 \\
\hline
\end{tabular}\\
{\small
\begin{tabular}{@{}p{\columnwidth}@{}}
 (1) Bond (1997); 
 (2) Galama \etal, 1997;
 (3) Djorgovski \etal, 1997a;
 (4) Jaunsen \etal, 1997;
 (5) Djorgovski \etal, 1997b;
 (6) Djorgovski \etal, 1997c;
 (7) Groot \etal, 1997;
 (8) Mignoli \etal, 1997;
 (9) Garcia \etal, 1997;
 (10) Fruchter \etal, 1997.
\end{tabular}
}
\end{center}
\end{table}

This model not only predicts that the optical luminosity should
first increase and then decline, but it also reproduces correctly the
time--delay between the beginning of the afterglow and the onset of 
optical emission. If the emission is pure synchrotron, there
will be a minimum electron energy given approximately by $\gamma_m = 
m_p \gamma /m_e$ in the shell frame, emitting synchrotron 
photons at the turnover frequency $\nu_m = 3\gamma_m^2 e B/4\pi m_e c$, 
with $B$ given by the usual equipartition argument $\propto \gamma$. In the
Earth frame, this frequency is given by
\begin{equation}
\label{turnover}
\nu_m = \frac{3}{4\pi} \left(\frac{m_p}{m_e}\right)^2
\frac{e \sqrt{8\pi n_1 m_p c^2}}{m_e c} \gamma^4\;.
\end{equation}
As the shell decelerates, $\nu_m$ will enter the optical region, 
$\approx 2\; eV$. This occurs for $\gamma_{opt} = 3.6 n_1^{-1/8} $, 
independent of all burst parameters except for the very weak dependence
upon the density of the surrounding medium. From Eq. \ref{tempo}, 
$\gamma=\gamma_{opt}$ is reached at a time $t_{opt}$ given by 
$t_{opt}/t_\circ \approx (\eta/\gamma_{opt})^{(7-2\alpha)/(3-2\alpha)}$. 
Introducing numerical values, I find, for $\alpha = 2$, 
$t_{opt}= 3.4$~days, in agreement with the observed $t_{opt} = 2.5$
(Djorgovski \etal, 1997b).
In the case of {\it GRB 970228} instead, corresponding to
$\alpha = 0$, I find $t_{opt} = 3.9 h$, well before the detection of 
the optical transient (van Paradijs \etal, 1997), and thus in agreement 
with observations. 

Eq. \ref{turnover} is quite remarkable: not only does it show but 
a weak dependence upon the surrounding matter density and no other
dependence upon burst luminosity, distance or beaming angle, it also tells
us that, by mapping the times of flaring of the burst at different 
wavelengths, one can determine the time--dependence of the shell
expansion, Eq. \ref{tempo}. Thus the hydrodynamics of the problem at hand
is, at least potentially, directly amenable to testing.

It can easily be seen from Eq. \ref{turnover} that the turnover frequency
cannot possibly have entered the radio region within the $5d$ time lapse
within which the source was seen to flare (Frail \etal, 1997), so that it 
seems reasonable that this radio flaring is due to the source becoming 
optically thin. The optical depth below the turnover frequency for $\nu < 
\nu_m$, scales as $\tau = \tau_m (\nu/\nu_m)^2$, due to the presence of a
flat, most likely thermal, energy distribution of electrons (Tavani 1996).
The optical depth at the turnover frequency $\nu_m$ can be computed 
approximately from the high frequency limit ($\nu > \nu_m$) as 
\begin{equation}
\label{taum}
\tau_m = \frac{3.1\times 10^{-8}}{\eta^{20/3}} \left(\frac{t_E}{t_\circ}
\right)^{\frac{20}{3} \frac{3-\alpha}{7-2\alpha}}
\end{equation}
from which we find that the frequency which is becoming optically thin 
at any given time, $\nu_{ot} = (\tau_m)^{1/2} \nu_m$ is given by
\begin{equation}
\label{opticallythin}
\nu_{ot} =  2.5 \; G\!H\!z \left(\frac{t_E}{5\; d}\right)^{-\frac{2}{3}
\frac{3-\alpha}{7-2\alpha}}\;.
\end{equation}
The scaling has been chosen so to show that the detection of Frail \etal\/
(1997), at $8.46\; GHz$ a mere $5\; d$ after the burst is correctly
reproduced. Also, the corresponding expected flux level, $F_{ot} = 
F_{\nu_m} (\nu_{ot}/\nu_m) ^{1/3}$, is given by 
\begin{equation}
\label{radioflux}
F_{ot} = 0.6 \; mJy \left(\frac{t_E}{5\; d}\right)^{
\frac{4\alpha-1}{3(7-2\alpha)}}\;,
\end{equation}
fortuitously close to the observations of Frail \etal \/ 1997, of
$0.43\; mJy$. It should also be noticed that the inverted spectrum
observed, $\propto \nu^{1.1}$, which is so important in ruling out
the hypothesis that the burst be due to a blazar, is characteristic
of a radio source caught in the process of becoming optically thin.

\section{Discussion and summary}

An alternative and comprehensive model accounting for several properties
of the afterglow of {\it GRB 970508} has been presented by Waxman 
(1997b), who postulates that the expansion is adiabatic. This makes
the observed radiation small with respect to the afterglow energy,
with the balance of this energy going into adiabatic losses; in fact,
he postulates large ($E\approx 10^{52}\; erg$) energy releases. There are
two observational major differences between his model and the present one. 
First, in Waxman's model, the slow decrease of the X--ray luminosity 
cannot be accounted for: in fact, Inverse Compton scattering (Waxman 1997a)
is incapable of decreasing the number of photons (and thus the X--ray 
luminosity) produced by synchrotron radiation, because it is well--known
that the total optical depth to Thompson scattering in GRB ejecta shell
is very small, $\approx 10^{-6}$ (Sari, Narayan and Piran 1996). Thus 
the model should show the same time--dependence of the X--ray luminosity
as {\it GRb 970228}, which it does not (Piro \etal, 1997). Second,
since in his model the cooling time is long compared to the local dynamical
time--scale, Waxman (1997b) expects an X--ray spectrum going as $\nu
^{-(p-1)/2}$, while I expect $\nu^{-p/2}$, exactly like in the burst proper.
Thus spectral observations will be able to tell whether the expansion
is adiabatic (the X--ray spectrum is harder in the afterglow, Waxman 1997b),
or highly radiative (afterglow and burst having similar spectrum, this paper).

The main results of this paper are as follows:
\begin{itemize}
\item The X--ray afterglow luminosities of both {\it GRB 970228} 
and {\it GRB 970508} are well--fitted by the deceleration of
a radiative relativistic shell, plowing through external matter;
in the first case, a constant density allows a good fit, while in the
second one a power--like density distribution $\propto r^{-2}$,
like that left over by a prior mass loss episode, is required;
\item the delay ($\approx 30\; s$) between the first peak and the
onset of the afterglow, in the X and $\gamma$ emission of {\it GRB 970228}
is quantitatively explained by assuming that the first peak results from an
internal shock, and the afterglow from an external shock;
\item the existence of a delay between the optical luminosity maximum 
and the $\gamma$ peak in {\it GRB 970508} can most easily be explained by
expansion in a nonuniform external medium, thus strengthening the 
interpretation of the X--ray light curve;
\item the optical time--delay for {\it GRB 970508} is well reproduced
as $3.4$~days;
\item the late appearance of the radio flux, the observed flux level, 
and the peculiar inverted spectrum are all easily accounted for in this model.
\end{itemize}

\vskip 2truecm
Thanks are due to Eli Waxman for showing me his work well before
publication, to G. Ghisellini, G.C. Perola and E. Waxman for helpful 
scientific discussions, and to F. Pacini for organizing the Elba
Workshop on Gamma Ray Bursts where this paper was completed.

\begin{figure}
\caption \/ X--ray flux in the afterglow of {\it GRB 970228}; data points
are from Costa \etal, 1997a; the theoretical curve is from Eq. \ref{xflux}. 

\caption \/ Optical observations of the afterglow of {\it GRB 970508};
data points are from Table I; straight lines (to be compared with the 
predictions of Eq. \ref{optflux}) are eye--fit to the data.
\end{figure}

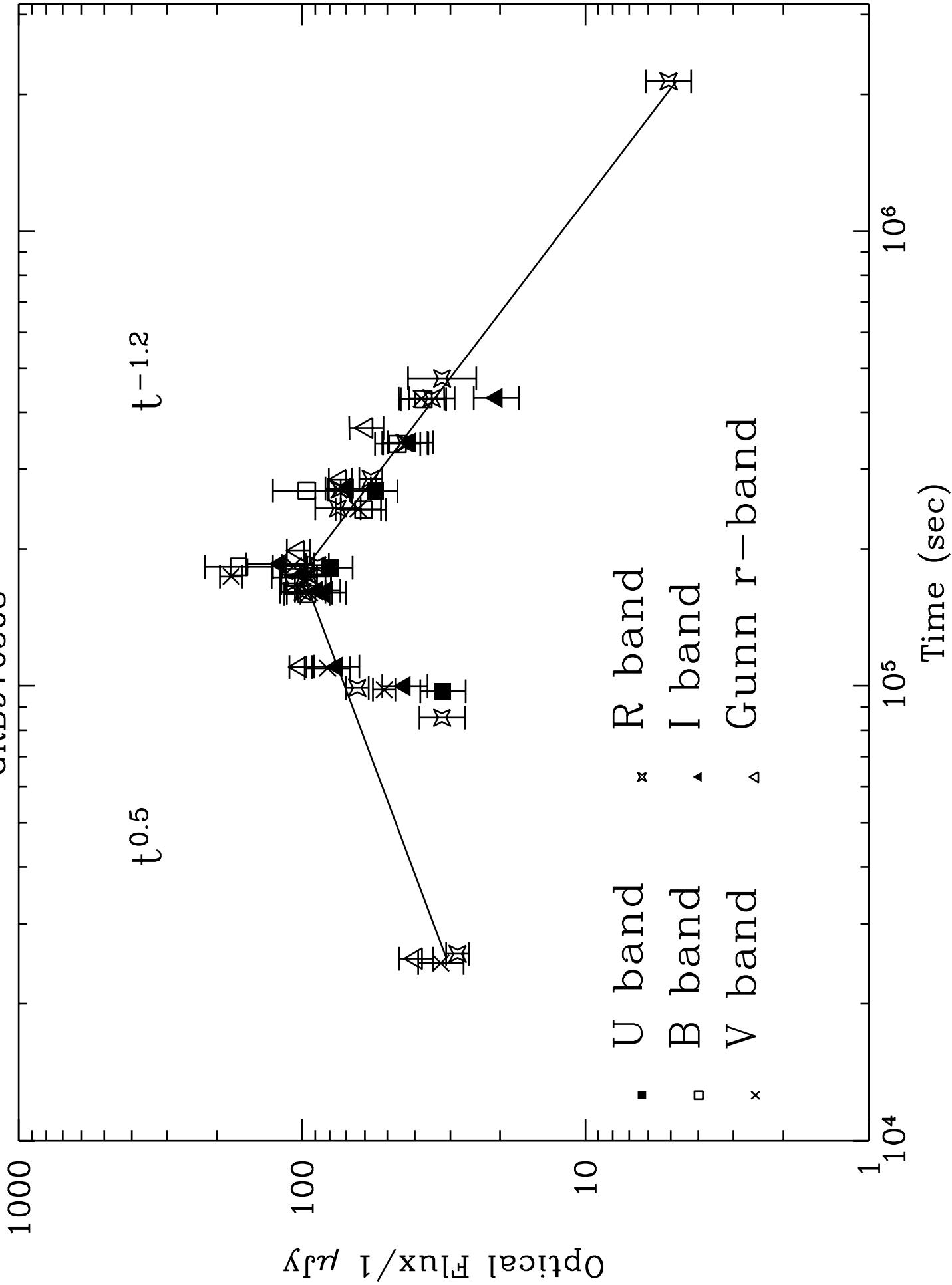
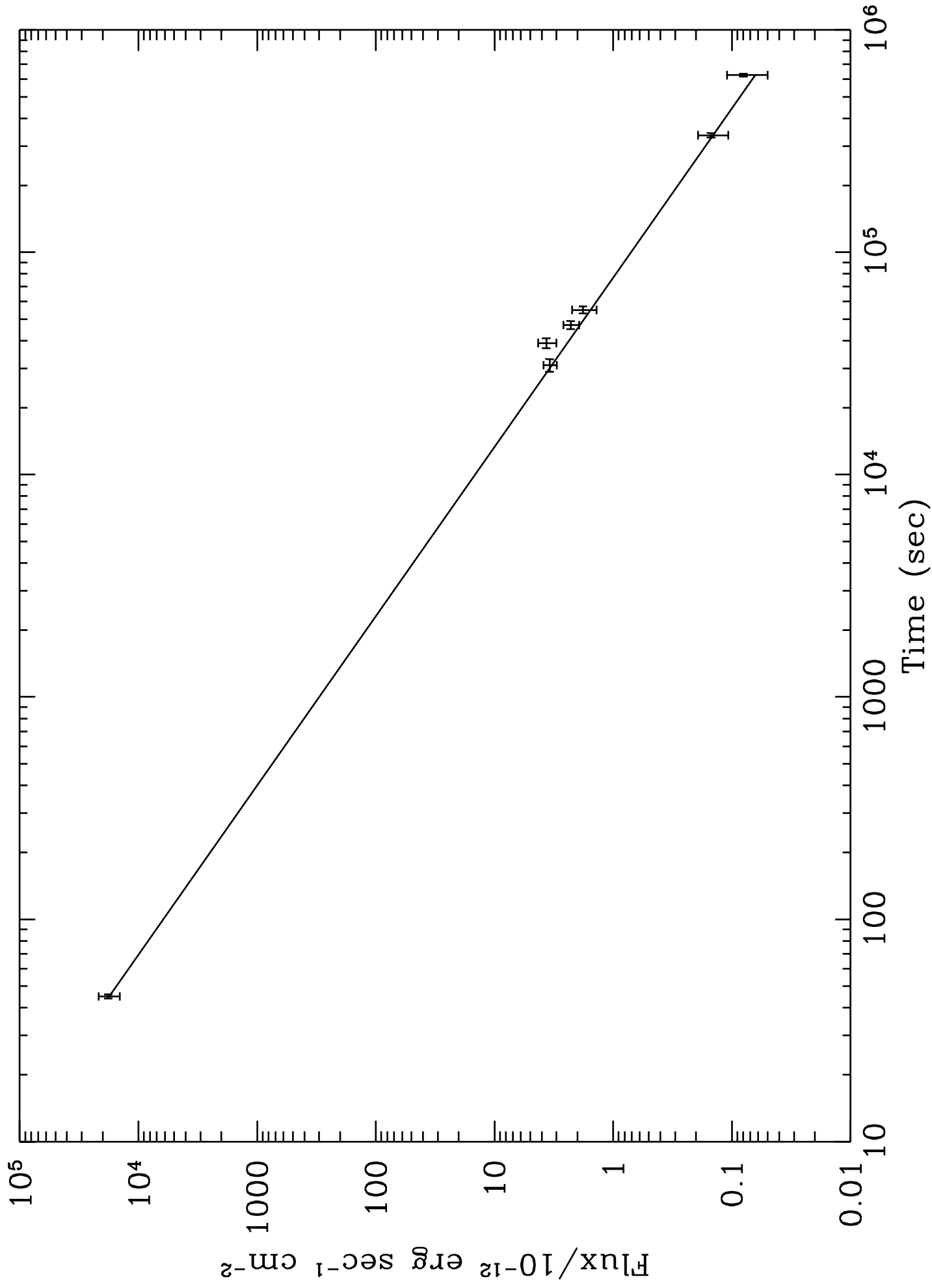
\end{document}